\documentclass[twocolumn]{aastex62}

\usepackage{graphicx,amsmath,lipsum}

\received{\today}
\revised{...}
\accepted{...}
\submitjournal{ApJL}

\shorttitle{Multi-wavelength QPP in a stellar superflare}
\shortauthors{Kolotkov et al.}

\begin{document}

\title{Multi-wavelength quasi-periodic pulsations in a stellar superflare}

\correspondingauthor{Dmitrii Y. Kolotkov}
\email{D.Kolotkov.1@warwick.ac.uk}

\author{Dmitrii~Y. Kolotkov}
\affiliation{Centre for Fusion, Space and Astrophysics, Department of Physics, University of Warwick, CV4 7AL, UK}
\affiliation{Institute of Solar-Terrestrial Physics, Irkutsk 664033, Russia}

\author{Valery~M. Nakariakov}
\affiliation{Centre for Fusion, Space and Astrophysics, Department of Physics, University of Warwick, CV4 7AL, UK}
\affiliation{St. Petersburg Branch, Special Astrophysical Observatory, Russian Academy of Sciences, 196140, St. Petersburg, Russia}

\author{Robin Holt}
\affiliation{Centre for Fusion, Space and Astrophysics, Department of Physics, University of Warwick, CV4 7AL, UK}

\author{Alexey~A. Kuznetsov}
\affiliation{Institute of Solar-Terrestrial Physics, Irkutsk 664033, Russia}

\begin{abstract}
We present the first multi-wavelength simultaneous detection of QPP in a superflare (more than a thousand times stronger than known solar flares) on a cool star, in soft X-rays (SXR, with XMM-Newton) and white light (WL, with Kepler). It allowed for the first-ever analysis of oscillatory processes in a stellar flare simultaneously in thermal and non-thermal emissions, conventionally considered to come from the corona and chromosphere of the star, respectively. The observed QPP have {periods $1.5 \pm 0.15$ hours (SXR) and $3 \pm 0.6$ hours (WL)}, and correlate well with each other. The unique relationship between the observed parameters of QPP in SXR and WL allowed us to link them with oscillations of the electric current in the flare loop, which directly affect the dynamics of non-thermal electrons and indirectly (via Ohmic heating) the thermal plasma. These findings could be considered in favour of the equivalent LCR-contour model of a flare loop{, at least in the extreme conditions of a stellar superflare.}
\end{abstract}


\section{Introduction} \label{sec:intro}

Solar flares and coronal mass ejections are the most powerful physical phenomena in the solar system, and the key driver of space weather \citep[][]{2011ASPC..448..231S, 2017LRSP...14....2B}. Physical processes operating in flaring sites, such as magnetic reconnection, charged particle acceleration and turbulence remain key challenges of plasma astrophysics \citep[e.g.,][]{2011LRSP....8....6S}. The parametric range of the flare research is significantly broadened by observing flares on other stars, including those of solar type. In particular, observations of stellar superflares, with the released energy several orders of magnitude higher than in the most energetic observed solar flare \citep[e.g.,][]{2012Natur.485..478M} are important for assessing whether the Sun is capable of producing a devastating solar superflare.

{Despite a tremendous effort in understanding the physics of flares, revealing a comprehensive generic model of a flare, consistent with observations remains one of the longest-standing and impactful questions for space weather research. In particular, the so-called \textit{standard} model of a solar flare \citep{2011LRSP....8....6S}, based on a magnetohydrodynamic (MHD) description of the processes of current-sheet development, magnetic field restructuring, and subsequent acceleration of charged particles, captures well the global picture of a flare, but struggles with explaining more specific questions such as what triggers the flare, how the released energy is split between different channels (including manifestations in specific electromagnetic bands), what determines the characteristic timescales, etc. Considering the inductive property of solar atmospheric plasma configurations (i.e. the induction of the magnetic field by electric current systems and vice versa) as a fundamental storage of free magnetic energy, \cite{1967SoPh....1..220A} proposed a flare model based on the analogy with a closed electric circuit. In this model, the disruption of the electric current leads to the explosive release of the whole magnetic energy of the circuit via local Ohmic heating (cf. switches in high-power transmission networks).}

An intriguing process not predicted by the standard flare model but commonly observed in solar flares are quasi-periodic pulsations (QPP) of the emitted radiation, which have been detected in all spectral bands \citep[see e.g.,][]{2009SSRv..149..119N, 2020STP.....6a...3K}. Typically, QPP appear as subsequent increases and decreases in the emission intensity, usually lasting for several cycles only. Often, oscillatory patterns in QPP are non-stationary, i.e., are subject to amplitude and period modulations \citep[see e.g.,][]{2019PPCF...61a4024N}.
Physical mechanisms responsible for QPP are subject to intensive ongoing studies. It is expected that these mechanisms could be divided into three main groups: the modulation of the emitting plasma by MHD oscillations, repetitive magnetic reconnection which is periodically induced by an MHD oscillation, and spontaneous repetitive reconnection \citep[see][for a comprehensive review]{2018SSRv..214...45M}.
{Taking the effective inductance ($L$), capacitance ($C$), and resistance ($R$) of coronal plasma configurations into account in the Alfv\'en's flare model \citep{1967SoPh....1..220A}, oscillatory variations of the electric current in the flare loop (considered as an equivalent LCR-contour)  were theoretically predicted as a unique feature of the model, naturally leading to QPP \citep{1998A&A...337..887Z, 2009SSRv..149...83K, 2021SSRv..217...66Z}. In general,} the identification of the mechanism for a QPP requires simultaneous observations at different wavelengths which are associated with thermal and non-thermal emission, i.e., in different spectral bands, providing the crucial information about the release and transport of the flare energy through different layers of the solar atmosphere \citep{2021SSRv..217...66Z}.
{In other words, no contemporary flare model is acceptable unless it adequately accounts for the phenomenon of QPP.}

QPP are detected in stellar flares too, including the radio \citep[e.g.,][]{2001A&A...374.1072S},  soft X-rays (SXR) \citep[e.g.,][]{2005A&A...436.1041M}, UV \citep[e.g.,][]{2018MNRAS.475.2842D}, and white light (WL) \citep[e.g.,][]{2016MNRAS.459.3659P} bands. Empirical properties of QPP in solar and stellar flares have similarities \citep{2016ApJ...830..110C} which could indicate the analogy in underlying physical processes. 
In particular, QPP in stellar flares have been detected at different wavelengths within the same spectral band: in optics \citep{2000A&A...364..641Z}, in radio \citep{2001A&A...374.1072S}, and in SXR \citep{2019A&A...629A.147B}. \citet{2019A&A...622A.210G} analysed simultaneous observations of stellar superflares in the WL (with Kepler) and SXR (with XMM-Newton), and found 500-s QPP in an SXR flare lightcurve on the M2 class star HCG 273. However, the corresponding periodicity was not revealed in the WL band. Thus, to the best of our knowledge there have been no simultaneous observations of QPP in stellar superflares in distinctly different electromagnetic wavebands which would allow for distinguishing between thermal and non-thermal flare emissions so far.


The omnipresence of various transient wave and oscillatory phenomena in elastic media such as solar and stellar atmospheres naturally allows for the use of them as a unique tool for seismological diagnostics of the local plasma conditions and processes, which could not be measured otherwise. The successful application and potential of this approach has been confidently demonstrated for studying the local plasma conditions in the corona of the Sun known as coronal MHD seismology \citep[see][for the most recent review]{2020ARA&A..58..441N}, for which direct spatially and temporally resolved observations of MHD waves and oscillations are ubiquitously available. For stars, in the absence of direct spatially-resolved observations, the only proxy of wave dynamics in their atmospheres is time-resolved observations of QPP in the lightcurves of stellar flares and superflares. In other words, the phenomenon of QPP offers a unique but yet unexplored source of information about ongoing processes and physical conditions in atmospheres of stars, through the transfer of the method of MHD coronal seismology from solar physics to the realm of stellar physics and exploitation of the solar-stellar analogy.
{Such a promising perspective of a QPP-based \emph{stellar MHD coronal seismology} clearly justifies the interest and high demand in multi-wavelength observations of QPP in stellar flares in distinctly different spectral bands. This would enable, in particular, studies of the development of flare energy releases at different layers of stellar atmospheres, and allow for advancing of our understanding of the physics of flares and solar-stellar analogy, in general.} 

In this Letter, we present the first-ever detection of QPP signals in a stellar superflare simultaneously in the SXR and WL bands, typically associated with thermal and non-thermal emissions from the corona and chromosphere of the star, respectively. We identify a specific relationship between the observed WL and SXR QPP parameters (ratio of the oscillation periods), which could be interpreted as a {plausible}  feature of the equivalent LCR-contour oscillations in a stellar superflare {\citep[among at least fifteen physical mechanisms/models that have been proposed to explain QPP in flares, see][]{2021SSRv..217...66Z, 2020STP.....6a...3K, 2018SSRv..214...45M}.}



\section{Observations} \label{sec:observations}
We use simultaneous observations of a superflare occurred on 30 November 2009 on the M3V class star KIC 8093473, obtained in the white light (WL) and soft X-ray (SXR) bands with the Kepler and XMM-Newton (0.2--12\,keV, with the EPIC PN detector) missions, respectively (see the left-hand panel in Fig.~\ref{fig:lightcurve_fourier}).
{By the peak X-ray flux, the analysed stellar superflare was previously shown to be equivalent to X14,700 GOES-class solar flare \citep[][cf. the strongest X28-class solar flare ever observed]{2021ApJ...912...81K}.
{The analysed flare was selected by inspecting the Kepler \citep{2010Sci...327..977B} and XMM-Newton \citep{2001A&A...365L...1J, 2016A&A...590A...1R} databases. More specifically, among 69 stars observed simultaneously by both instruments \citep{2019A&A...628A..41P}, \citet{2021ApJ...912...81K} identified three stars with several flares (some of which likely consisting of multiple emission peaks) manifested simultaneously in the WL and X-ray bands. We study the presence of QPP in one of those complex flaring events; the visual inspection of the lightcurves of other flares did not show signatures of quasi-periodic variability and therefore they are not discussed here.}

{The host} star is considered to be a young tidally locked binary consisting of two more-or-less similar red dwarfs \citep{2021ApJ...912...81K}, with the orbital and rotation periods of 6.043 days \citep{2014ApJS..211...24M} and temperature of 3400 K \citep{2018yCat.1345....0G}, at a distance of 206 pc \citep{2018yCat.1345....0G}; the analysed flare likely occurred on one of the components of the binary rather than in the interstellar space \citep{2021ApJ...912...81K}.
The flare showed a rapid increase in the SXR {flux}, developing into a well pronounced flare peak {followed by several weaker quasi-periodic peaks in the flare decay phase}.
In the WL observations, the event is manifested as a series of consecutive {quasi-periodic peaks} superposed on the background stellar irradiance modulated by the rotation period.

The WL observations by Kepler have time cadence of {29.5\,min}{, i.e. the Kepler data represent the WL flux binned (averaged) over this time interval}. The SXR observations from XMM-Newton used in this work have bin size (effective cadence) of 4.5\,min.
Based on the solar-stellar analogy, the emission mechanism for the WL radiation in flares is associated with the blackbody radiation from 
lower layers of the stellar atmosphere, heated by non-thermal electrons \citep{2010ARA&A..48..241B}. In turn, the emission mechanism for the SXR radiation is associated with thermal radiation from a hot ($\sim 30$\,MK) coronal plasma \citep{2004A&ARv..12...71G}.
{Earlier, \citet{2021ApJ...912...81K} rigorously demonstrated that the largest SXR flare peak in the analysed flare is delayed with respect to its WL counterpart by $15\pm 23$\,min (with the uncertainty determined mainly by the time resolution of Kepler), which suggests the presence of the Neupert effect \citep{1968ApJ...153L..59N} as a characteristic signature of flare thermal and non-thermal emissions. In addition, the ratio of the WL to X-ray peak amplitudes was found to increase gradually with time, that was interpreted as a manifestation of the soft-hard-harder evolution of non-thermal electrons in the flare \citep[see e.g.][]{2011SSRv..159...19F}.}
{At the same time, we would note that, despite being very common and apparently present in the considered event, the Neupert effect is not observed in all solar and stellar flares.}

\begin{figure*}
	\centering
	\includegraphics[width=\linewidth]{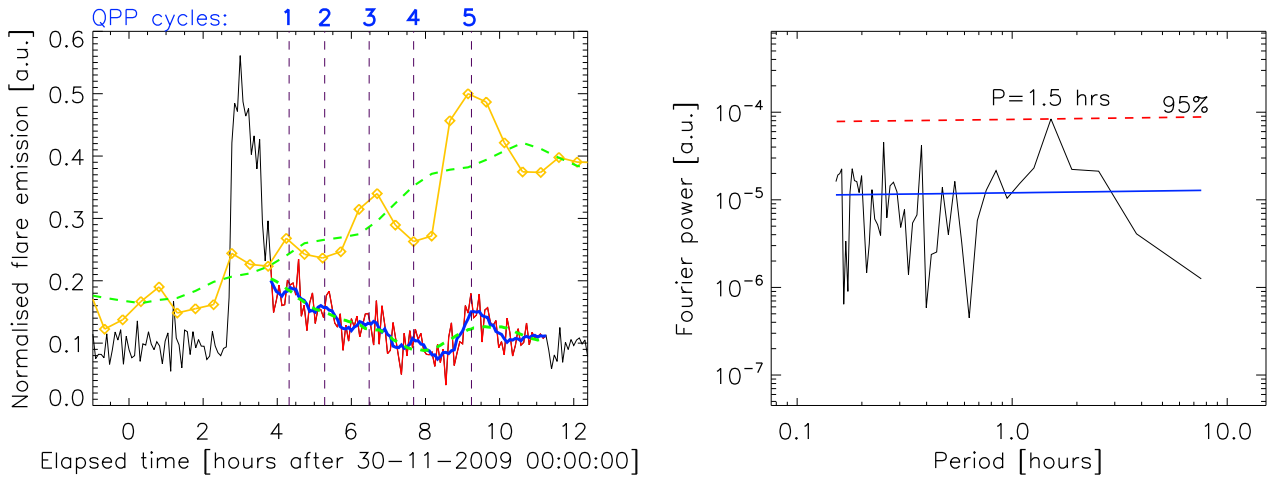}
	\caption{Left: Normalised lightcurves of a superflare event, analysed in this work, as seen by XMM-Newton in the soft X-ray band (SXR, the black--red solid curve) and by Kepler in the white-light band (WL, the yellow line).
		The gradual trend of the WL lightcurve is caused by the background WL irradiance modulated by the star's rotation {(with a period of 6.043 days, see Sec.~\ref{sec:observations})}.
		The red curve shows the SXR signal in the time interval of interest, during which signatures of QPP are detected. The green dashed lines show the long-term trends of the WL and SXR signals.
		The blue solid line shows the SXR signal's interval of interest smoothed over 30\,min thus mimicking the time resolution of the WL lightcurve (used in Fig.~\ref{fig:cross-corr}). Both lightcurves are normalised to some arbitrary constants for better visualisation. Right: The Fourier power spectrum of the SXR signal of interest shown in red in the left-hand panel with the corresponding long-term trend subtracted. The blue solid line shows the best-fit of the spectrum by a power-law function. The red dashed line indicates the statistical significance level of 95\%, estimated as described in Sec.~\ref{sec:qpp}.
	}
	\label{fig:lightcurve_fourier}
\end{figure*}

\section{QPP detection and analysis} \label{sec:qpp}

{As the phenomenon of QPP is traditionally distinguished from the flux variability caused by the flare itself \citep[see e.g.][]{2016ApJ...830..110C},} for the detection of QPP in this work, we focus on the decay phase of the SXR flare lightcurve{, i.e. after the main flare peak} (see the interval of interest shown in red in the left-hand panel of Fig.~\ref{fig:lightcurve_fourier}).
Likewise, we use the Fourier transform-based method complemented by a rigorous assessment of the statistical significance of the obtained Fourier components in comparison with power-law distributed noisy background \citep{2005A&A...431..391V, 2017A&A...602A..47P}, as the most robust and straightforward approach for detection of QPP in solar and stellar flares \citep{2019ApJS..244...44B}.
A slowly-varying trend of the original signal, $T_\mathrm{SXR}(t)$ is subtracted before applying the Fourier analysis.
In this work, the SXR trend is obtained by smoothing the original signal over 150\,min, using a Savitzky--Golay polynomial filter.
Having the presence of QPP in the SXR flare lightcurve assessed, we calculate the SXR modulation depth as $\delta F_\mathrm{SXR}(t) = [F_\mathrm{SXR}(t) - T_\mathrm{SXR}(t)]/T_\mathrm{SXR}(t)$.

For the WL emission, we obtain a slowly-varying trend $T_\mathrm{WL}(t)$ by smoothing the original signal over 265\,min, which gives the modulation depth $\delta F_\mathrm{WL}(t) = [F_\mathrm{WL}(t) - T_\mathrm{WL}(t)]/T_\mathrm{WL}(t)$. Due to low time resolution of the WL signal, we assess the presence of quasi-periodic behaviour in it via cross-correlating $\delta F_\mathrm{WL}(t)$ with $\delta F_\mathrm{SXR}(t)$ and $[\delta {F}_\mathrm{WL}(t)]^2$ with $\delta F_\mathrm{SXR}(t)$.

For checking the cross-correlation between the latter, i.e. $[\delta {F}_\mathrm{WL}(t)]^2$ and $\delta F_\mathrm{SXR}(t)$, we represent $\delta F_\mathrm{WL}(t)$ as $\delta F_\mathrm{WL}(t) = A_0(t)\cos\omega t$ where $\omega$ and $A_0(t)$ are the characteristic oscillation frequency and instantaneous amplitude determined as
\begin{equation}\label{eq:instant_amp}
	A_0(t) = |\delta F_\mathrm{WL}(t) + i {\cal H}\left\{\delta F_\mathrm{WL}(t)\right\} |,
\end{equation}
where ${\cal H}\left\{\delta F_\mathrm{WL}(t)\right\}$ is the Hilbert transform of $\delta F_\mathrm{WL}(t)$ \citep{1998RSPSA.454..903H}. Applying the trigonometric identity $\cos^2\alpha = (1+\cos 2\alpha)/2$, $\delta F_\mathrm{WL}^2(t)$ can be re-written as $\delta F_\mathrm{WL}^2(t) = A_0^2(t)[1+\cos 2\omega t]/2$. From here, we isolate $\delta \tilde{F}_\mathrm{WL}^2(t)\equiv A_0(t)\cos 2\omega t$ which becomes
\begin{equation}\label{eq:deltaf2}
	\delta \tilde{F}_\mathrm{WL}^2(t) = \frac{2[\delta {F}_\mathrm{WL}(t)]^2 - A_0^2(t)}{A_0(t)}.
\end{equation}
The new signal $\delta \tilde{F}_\mathrm{WL}^2(t)$ takes into account a non-stationary oscillation amplitude $A_0(t)$ of the input signal $\delta {F}_\mathrm{WL}(t)$ and has approximately zero mean, that allows for direct comparison of $\delta \tilde{F}_\mathrm{WL}^2(t)$ with $\delta F_\mathrm{SXR}(t)$ (see the right-hand panel in Fig.~\ref{fig:qpp_both}).

\section{Results}
\label{sec:results}
The Fourier power spectrum in Fig.~\ref{fig:lightcurve_fourier} clearly shows the presence of QPP with period of about 1.5 hours in the detrended SXR flux in the decay phase of the flare, comprised of five well-pronounced oscillation cycles in the time domain (see the left-hand panel of Fig.~\ref{fig:lightcurve_fourier}).

The star's WL emission during the time interval of interest, for which QPP are detected in SXR, also exhibits a quasi-periodic variation of intensity. Moreover, the QPP cycles 1, 3, 5 (2, 4) in the SXR band are seen to be in-phase (anti-phase) with the WL pulsations.
The observed tendency strongly suggests that the period of modulation of the WL emission is two times longer than that in the SXR signal, thus being about 3 hours.
This observational finding provides the first-ever simultaneous detection of QPP signatures in a stellar superflare in two distinctly different wavebands, which could be associated with thermal and non-thermal emission mechanisms, and their comparative analysis.

\begin{figure*}
	\centering
	\includegraphics[width=\linewidth]{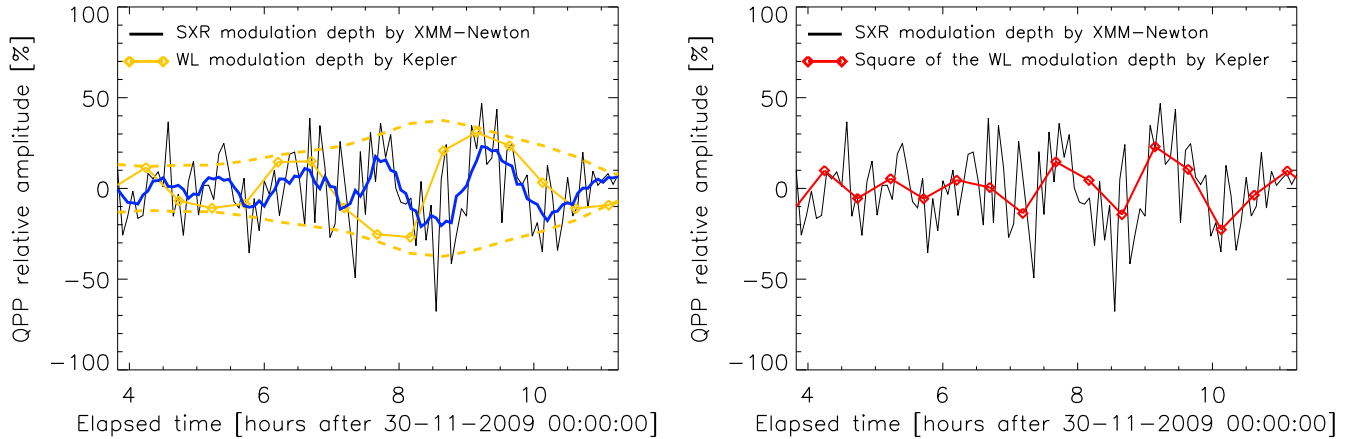}
	\caption{Left: The modulation depth, that is ratio of the detrended original flare lightcurve to its long-term trend, of QPP observed in the SXR band, $\delta F_\mathrm{SXR}(t)$, and in the WL band, $\delta F_\mathrm{WL}(t)$.
		The blue solid line is analogous to that shown in Fig.~\ref{fig:lightcurve_fourier}.
		The yellow dashed lines show the instantaneous amplitude of $\delta F_\mathrm{WL}(t)$, obtained by Eq.~(\ref{eq:instant_amp}). Right: The SXR modulation depth $\delta F_\mathrm{SXR}(t)$ (the same as shown in the left-hand panel) and a normalised square of the WL modulation depth $\delta \tilde{F}_\mathrm{WL}^2(t)$ determined by Eq.~(\ref{eq:deltaf2}).
	}
	\label{fig:qpp_both}
\end{figure*}

{As there are only a few peaks with a few data points per period for both WL and SXR signals, we assess the uncertainties in the WL and SXR period measurements, $P_\mathrm{WL} \pm \Delta_\mathrm{WL}$ and $P_\mathrm{SXR} \pm \Delta_\mathrm{SXR}$ by a half-width of the FFT frequency bin for the considered duration of the time interval of interest, $\Delta f \approx 0.07$ hours$^{-1}$. Using this, $\Delta_\mathrm{WL}$ and $\Delta_\mathrm{SXR}$ can be estimated as $\Delta_\mathrm{WL} = P_\mathrm{WL}^2 \Delta f$ and $\Delta_\mathrm{SXR} = P_\mathrm{SXR}^2 \Delta f$, which gives $P_\mathrm{SXR} = 1.5 \pm 0.15$ hours and $P_\mathrm{WL} = 3 \pm 0.6$ hours. The uncertainty in the period ratio $P_\mathrm{WL}/P_\mathrm{SXR}$ then becomes $(P_\mathrm{WL}/P_\mathrm{SXR}) \sqrt{(\Delta_\mathrm{WL}/P_\mathrm{WL})^2 + (\Delta_\mathrm{SXR}/P_\mathrm{SXR})^2}$, which gives $P_\mathrm{WL}/P_\mathrm{SXR} = 2 \pm 0.4$.}

The doubling of the QPP period in the WL signal, detected for the original flare lightcurves (Fig.~\ref{fig:lightcurve_fourier}), is also evident in Fig.~\ref{fig:qpp_both} from comparison of the time history of the corresponding modulation depth signals, $\delta F_\mathrm{SXR}(t)$ and $\delta F_\mathrm{WL}(t)$.
To perform a rigorous cross-correlation analysis between the oscillatory patterns seen in SXR and WL, we use $\delta \tilde{F}_\mathrm{WL}^2(t)$ given by Eq.~(\ref{eq:deltaf2}) and $\delta F_\mathrm{SXR}(t)$.
The results of the cross-correlation analysis between $\delta \tilde{F}_\mathrm{WL}^2(t)$ and $\delta F_\mathrm{SXR}(t)$ are shown in Fig.~\ref{fig:cross-corr}, demonstrating the highest value of the cross-correlation coefficient {reaches 0.5} at zero time lag. We note that this estimate of the cross-correlation coefficient is strongly affected by noise present in the SXR observations. However, if we smooth the SXR signal over 30\,min, thus effectively reproducing the time resolution of Kepler observations, the value of the cross-correlation coefficient between $\delta \tilde{F}_\mathrm{WL}^2(t)$ and $\delta F_\mathrm{SXR}(t)$ at zero time lag increases to 0.82. The latter rigorously confirms strong correlation between the analysed multi-band stellar QPP signals, and supports the conclusion that the QPP oscillation period observed in WL is two times longer than that in SXR.

{To demonstrate that the observed correlation between the WL and SXR signals is not of a random nature, we performed the Fisher randomisation test \citep[see e.g.][for the application of Fisher randomisation to oscillations in sunspots]{2010A&A...513A..27C}. Our test implies the estimation of the probability of obtaining the cross-correlation coefficient between $\delta \tilde{F}_\mathrm{WL}^2(t)$ and $\delta F_\mathrm{SXR}(t)$ higher than 0.8 (at any time lag) after a random permutation of the data points in $\delta \tilde{F}_\mathrm{WL}^2(t)$. We found that the cross-correlation $> 0.8$ appears in less than 1\% of all $10^6$ random permutations considered, which indicates in favour of a non-random nature of the observed high correlation between QPP signals in WL and SXR with the confidence exceeding 99\%.}

\begin{figure}
	\centering
	\includegraphics[width=\linewidth]{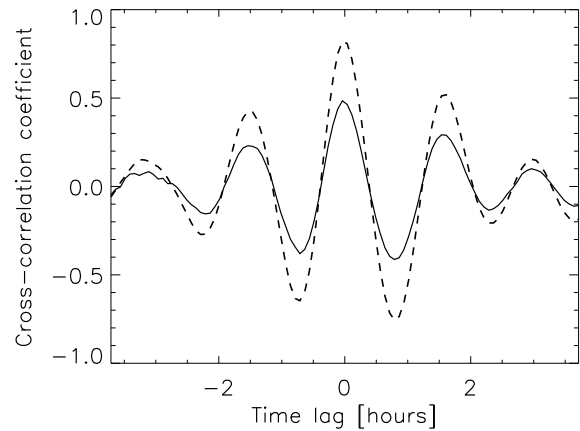}
	\caption{Results of the cross-correlation analysis between the SXR modulation depth $\delta F_\mathrm{SXR}(t)$ and the normalised square of the WL modulation depth $\delta \tilde{F}_\mathrm{WL}^2(t)$. The solid and dashed curves show the cross-correlation coefficients for the original $\delta \tilde{F}_\mathrm{WL}^2(t)$ and original $\delta F_\mathrm{SXR}(t)$, shown in the right-hand panel of Fig.~\ref{fig:qpp_both}, and for the original $\delta \tilde{F}_\mathrm{WL}^2(t)$ and $\delta F_\mathrm{SXR}(t)$ smoothed over 30\,min (thus effectively mimicking the time resolution of the WL signal, see Figs.~\ref{fig:lightcurve_fourier} and \ref{fig:qpp_both}), respectively.
	}
	\label{fig:cross-corr}
\end{figure}

\section{Discussion}
\label{sec:discuss}

The analysis reveals that QPP are present in both SXR and WL bands, with the oscillation {periods $1.5 \pm 0.15$\,hours and $3 \pm 0.6$\,hours}, respectively.
These periods are much shorter than the stellar rotation period ($\sim$6 days) and hence the observed oscillations cannot be attributed to the effect of rotational modulation \citep[cf. the events presented by][]{2021MNRAS.507.1723I}.	
{Likewise, the detection of similar QPP patterns in multi-instrumental independent observations is, in general, strong indication of their stellar origin \citep[see e.g.][]{2011A&A...530A..47I}.}
By the order of magnitude, the oscillation periods of WL QPP are consistent with the values detected in WL on other stars by e.g. \citet[][]{2013ApJ...773..156A, 2015ApJ...813L...5P, 2016MNRAS.459.3659P}. However, for the first time, a similar QPP pattern is simultaneously detected in SXR too. Moreover, the WL QPP has an oscillation period two-times longer than the SXR QPP.
This finding provides us with a crucial information for revealing the mechanism responsible for the oscillatory modulation of the emissions.
In the following we assume the widely accepted association of the SXR emission with the thermal emission from the flaring coronal loop \citep{2004A&ARv..12...71G}, and the WL emission with non-thermal emission from the loop's footpoints \citep{2010ARA&A..48..241B}.

Despite the variety of known mechanisms which could produce QPP \citep{2018SSRv..214...45M, 2021SSRv..217...66Z}, it is non-trivial to identify one which would explain why the SXR emission has an oscillation period two times shorter than of the WL emission. One possible option is that the flaring loop experiences the second parallel harmonic of the sausage oscillation \citep[e.g.][]{2020SSRv..216..136L}. The periodic narrowing of the magnetic flux tube which forms the coronal part of the loop, occurs alternatively in the opposite legs of the loop, causing the periodic precipitation of non-thermal electrons in the alternate footpoints due to periodically varying magnetic mirror ratio \citep{2009SSRv..149..119N}. As the characteristic loop length is comparable to the radius of the star \citep{2021ApJ...912...81K}, it is likely that only one footpoint appears to be on the visible hemisphere of the star. Thus, ${F}_\mathrm{WL}(t)$ comes from this footpoint only, with the oscillation period of the sausage mode, while ${F}_\mathrm{SXR}(t)$ comes from the coronal part with the oscillation period two times shorter. Sausage oscillation periods are 
\begin{equation}
	P_\mathrm{saus} \approx \frac{\pi r_\mathrm{min}}{2.4C_\mathrm{A0}},
\end{equation}
determined by the minor radius of the loop, $r_\mathrm{min}$ and the Alfv\'en speed $C_\mathrm{A0}$ inside it
\citep{2012ApJ...761..134N}. For $r_\mathrm{min} \approx 5$\,Mm and $C_\mathrm{A0}\approx 500$\,km\,s$^{-1}$, typical for solar flares, $P_\mathrm{saus} \approx 10$~s. 
For the periods of $10^4$\,s observed in this work, either $r_\mathrm{min}$ should be 1000 times larger than in the solar case, or 100 times larger with the Alfv\'en speed 10 times lower than in the solar case. Since the required radius of the magnetic loop, even for the lowest reasonable Alfv\'en speeds, becomes comparable to the loop length and the stellar radius, this explanation is unlikely.

Another possible interpretation of the observed QPP is provided by the equivalent LCR-contour model \citep{1998A&A...337..887Z, 2008PhyU...51.1123Z, 2009SSRv..149...83K}. In this model, the flaring active region is considered as a closed electric circuit \citep{1967SoPh....1..220A}. In the fully-ionised coronal part, the electric current is guided by the loop-like magnetic field. In the partly-ionised photosphere, the current can go across the field between the footpoints of the loop, closing the circuit. Such an electric circuit has a capacitance, inductance and resistance, determined by parameters of the plasma loop. Thus, the alternate electric current may experience oscillations with the period 
\begin{equation}
	P_\mathrm{LCR} \approx \frac{ (2\pi)^{3/2} \Lambda^{1/2}r_\mathrm{min}^2 \rho_0^{1/2}c }{ I_0} \left(1 + \frac{c^2 r_\mathrm{min}^2 B_{||0}^2}{4 I_0^2}\right)^{-1/2}, 
\end{equation}
where $\rho_0$, $I_0$ and $B_{||0}$  are the mass density, and the electric current and parallel magnetic field in the loop in the equilibrium, respectively; $c$ is the speed of light; and $\Lambda = \mathrm{ln}\frac{4 L}{\pi r_\mathrm{min}} - 7/4$, with $L$ being the loop length \citep[see, e.g.,][]{2009SSRv..149...83K, 2016ApJ...833..206T}.
For $L=700$\,Mm, $r_\mathrm{min}=30$\,Mm, $\rho_0 = 3\times10^{-12}$\,g\,cm$^{-3}$, $B_{||0} = 30$\,G, typical for stellar coronae \citep[see e.g.][]{1996ApJ...466..427M, 2005A&A...436.1041M, 2006A&A...456..323M, 2021ApJ...912...81K, 2021arXiv210810670R}, and a broad range of electric currents $I_0=10^8$--$10^{12}$\,A \citep[see e.g.][]{2009SSRv..149...83K}, we obtain $P$ about $10^4$~s. This value is consistent with the observed period.
The damping of such oscillations has been estimated to be rather weak \citep{1998A&A...337..887Z}, which is consistent with the observed behaviour too. In this scenario, the thermal emission from the coronal loop could be caused by the Ohmic dissipation of the current $I$, i.e., ${F}_\mathrm{SXR}(t) \propto I^2(t)$. Taking that $I(t) =  I_0 + \tilde{I}(t)$, where $\tilde{I}$ is the oscillating alternate current, we conclude that  ${F}_\mathrm{SXR}$ oscillates with double the period of $\tilde{I}(t)$ if the amplitude of $\tilde{I}$ is greater than $I_0$.

The presented results could be considered {in favour of} the equivalent LCR-contour nature of stellar flare loops.
{To the best of our knowledge, in the multitude of QPP models proposed hitherto \citep{2021SSRv..217...66Z}, there are no other mechanisms which could simultaneously a) cause modulation of thermal and non-thermal flare emissions, b) give observed periods of $10^4$\,s for reasonable combinations of stellar flare conditions, and c) explain the observed ratio of thermal and non-thermal QPP periods.}
The simultaneous oscillations in the thermal (SXR) and non-thermal (WL) emissions, and hence their similarities or differences, are hard to study in solar flares, for which the WL emission is rare and usually short-lived \citep{2017LRSP...14....2B}. On the other hand, extreme physical conditions in far more powerful stellar superflares make it possible to detect the WL emission co-existing with the soft X-ray emission for a sufficiently long time, and to study oscillatory processes superimposed.
Results obtained in this Letter open up a new opportunity for exploiting the analogy between solar and stellar flares via searching for similar correlations between oscillations in thermal and non-thermal flare emissions from the Sun, by using observations from existing and future high-resolution and high-sensitivity instruments.

{The intrinsic difficulties preventing the direct comparison and extrapolation of the results obtained in our work to solar flares are the lack of general understanding of differences and/or similarities between physics of WL flares on the Sun and other stars \citep{2010ARA&A..48..241B} and huge disparity in physical conditions and characteristic spatial and temporal scales in solar and stellar flares, which might lead to different observational manifestations of the same quasi-periodic modulation process.
One of the illustrations of these difficulties is the lack of observations of QPP in WL solar flares.}

{Our work suggests that the QPP period ratio of two in thermal and non-thermal emissions could be indicative of the operation of the LCR mechanism, at least in the extreme conditions of stellar superflares. On the other hand, the LCR model is not an exclusive mechanism for simultaneous QPP in thermal and non-thermal emissions. Indeed, there are at least several other physical mechanisms that could cause quasi-periodic modulation simultaneously in thermal and non-thermal bands, with their own unique observational features \citep{2021SSRv..217...66Z}. However, it is not clear whether one of those mechanisms could explain the observed period ratio.}

\acknowledgments
Analysis and interpretation of QPP in Sections 3--5 were supported by the Russian Science Foundation grant No. 21-12-00195.
D.Y.K. and V.M.N. acknowledge support from the STFC consolidated grant ST/T000252/1. V.M.N. was supported by the Russian Foundation for Basic Research Grant No.~18-29-21016. D.Y.K. and A.A.K. acknowledge support from the Ministry of Science and Higher Education of the Russian Federation.

\bibliographystyle{aasjournal} 



\end{document}